\documentclass{fizik}%
\usepackage{multicol,epsfig}

\vol{27}
\pyear{2003}
\received{\today}

\author[SAMUELSSON, SUKHORUKOV, B\"UTTIKER]{
\textbf{P. Samuelsson, E. Sukhorukov, and M. B\"uttiker}\\
\textit{D\'epartement de Physique Th\'eorique, Universit\'e de
Gen\`eve}\\
\textit{CH-1211 Gen\`eve 4, Switzerland}\\
}
\title{Orbital entanglement and violation of Bell inequalities in the
presence of dephasing}

\begin{document}
\maketitle

\begin{abstract}
We discuss orbital entanglement in mesoscopic conductors, focusing on
the effect of dephasing. The entanglement is detected via violation of
a Bell Inequality formulated in terms of zero-frequency current
correlations. Following closely the recent work by Samuelsson,
Sukhorukov and B\"uttiker\cite{Sam1}, we investigate how the dephasing
affects the possibility to violate the Bell Inequality and how system
parameters can be adjusted for optimal violation.

\end{abstract}

\section{Introduction}
Entanglement is one of the most intriguing features predicted by
quantum theory \cite{EPR,Bohr,Bohm}. It leads to correlations between
distant particles, which can not be described by any local, realistic
theory \cite{Bell}. This nonlocal property of entanglement has been
demonstrated convincingly in optics \cite{Gisin1,Weihs}, where
entangled pairs of photons have been studied over several decades
\cite{Zeilinger}. Apart from the fundamental aspects, there is a
growing interest in using the properties of entangled particles for
quantum cryptography \cite{GisinRMP} and quantum computation
\cite{SteaneRPP}.

Recently, much interest has been shown for entanglement of electrons
in solid state systems. A controlled generation and manipulation of
electronic entanglement is of importance for a large scale
implementation of quantum information and computation schemes.
Electrons are however, in contrast to photons, massive and
electrically charged particles, which raises new fundamental questions
and new experimental challenges. Most existing suggestions
\cite{Recher,Les,Burk00,Oliver,Recher02,Bena,Taddei02,Euges,Saraga,fazio,Sols}
are based on creating, manipulating and detecting spin-entangled pairs
of electrons. This requires experimental control of individual spins
via spin filters or locally directed magnetic fields on a mesoscopic
scale.

To overcome these difficulties, very recently spin-independent schemes
for creating and detecting orbital entanglement in a mesoscopic system
were proposed in normal-superconducting\cite{Sam1} as well as purely
normal systems\ \cite{Been,Sam2}. Energy-time entanglement
\cite{Gisin2} in a Franson like geometry \cite{Fran} provides another
spin-independent approach. In this paper, we discuss in detail the
properties of the orbital entanglement and how it can be detected via
violation of a Bell Inequality. In particular, we focus on the effect
of decoherence on the entanglement. We follow closely the discussion
in Ref.\ \cite{Sam1}, but the results are of relevance also for the
orbital entanglement studied in Refs.\ \cite{Been,Sam2}. An
investigation of the effect of decoherence on entanglement, discussing
the setup in Ref. \cite{Been}, is presented by van Velsen et al in
this issue.
  
\section{Entangled two-particle wavefunction.}

\begin{figure}[htb]
\begin{center}
\includegraphics[scale=0.5]{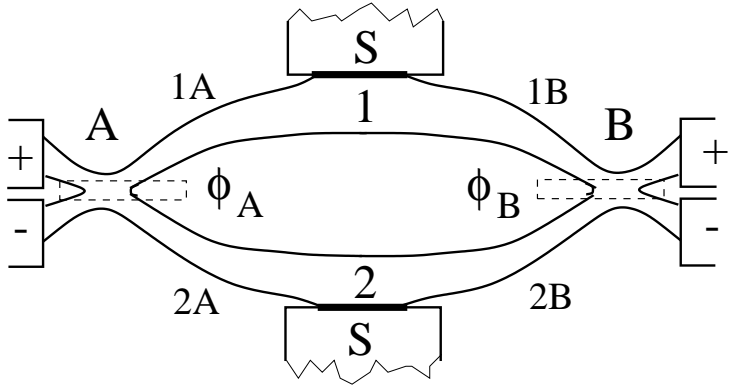}
\end{center}
\caption{The system of Ref. \cite{Sam1}: A single superconductor (S)
is connected to four normal arms via two tunnel barriers $1$ and $2$
(thick black lines). The arms are joined pairwise in beam splitters A
and B and end in normal reservoirs $+$ and $-$.}
\label{fig1}
\end{figure}

We consider the system shown in Fig. \ref{fig1}. A single
superconductor (S) [the upper and lower part in the figure connected
e.g. via loop] is weakly coupled to a normal conductor, a ballistic
two-dimensional electron gas, via two tunnel barriers $1$ and $2$ with
equal transparencies $\Gamma \ll 1$. The normal conductor consists of
four arms, $1A,1B,2A$ and $2B$, with equal lengths $L$. The arms $1A$
and $2A$ ($1B$ and $2B$) are crossed in a beam-splitter $A(B)$ and
then connected to normal reservoirs $+A$ and $-A$ ($+B$ and $-B$).
We consider the low temperature limit, $kT \ll eV$. A negative voltage
$-eV$ is applied to all the normal reservoirs and the superconductor
is grounded. The voltage $eV$ is smaller than the superconducting gap
$\Delta$, so no single particle transport takes place. It is further
assumed that the size of the system is smaller than the phase breaking
length.

The superconductor as a source of spin-entangled electrons has been
discussed in Refs.  \cite{Recher,Les,Bena,Sols}.  As shown in
Ref. \cite{Sam1}, due to the coherent properties of the
superconducting condensate, the superconductor can also act as an
emitter of orbitally entangled pairs of electrons. To lowest order in tunnel
barrier strength $\Gamma \ll 1$, a pair of electrons is emitted either
through contact 1 or contact 2, The state is thus a linear
superposition of emitted pairs, an orbitally entangled two-particle
state. In a second quantization language, the state of the entangled
pair can be written as \cite{Sam1}
\begin{eqnarray}
|\Psi_{out}\rangle=|\bar 0\rangle+|\tilde \Psi\rangle+|\Psi\rangle,
\label{entstate}
\end{eqnarray}
where the states
\begin{eqnarray}
|\tilde \Psi\rangle&=&\frac{i\Gamma}{4}\int_{0}^{eV}dE
 \sum_{j=1,2}\sum_{\eta=A,B} \left[c_{j\eta}^{\uparrow\dagger}(E)c_{j\eta}^{\downarrow\dagger}(-E)-c_{j\eta}^{\downarrow\dagger}(E)c_{j\eta}^{\uparrow\dagger}(-E) \right]|\bar 0\rangle, \nonumber \\
 |\Psi\rangle&=&\frac{i\Gamma}{4}\int_{-eV}^{eV}dE\left[c_{1A}^{\uparrow\dagger}(E)c_{1B}^{\downarrow\dagger}(-E)-c_{1A}^{\downarrow\dagger}(E)c_{1B}^{\uparrow\dagger}(-E)+c_{2A}^{\uparrow\dagger}(E)c_{2B}^{\downarrow\dagger}(-E)-c_{2A}^{\downarrow\dagger}(E)c_{2B}^{\uparrow\dagger}(-E)\right]|\bar 0\rangle.
\label{entNSwavefcn12}
\end{eqnarray}
Here, e.g. an operator $c^{\uparrow\dagger}_{1A}(E)$ describes the
creation of an electron with spin up at energy $E$, emitted through
contact $1$ and propagating towards beamsplitter $A$. The state $|\bar
0\rangle$ is the groundstate, a filled Fermi sea with electrons at all
energies $E<-eV$ and $|\Psi\rangle$ ($|\tilde \Psi\rangle$) describes
orbitally entangled wavepacket-like states with two electrons
propagating to different (the same) beamsplitters. As is shown in
Ref. \cite{Sam1}, the state $|\tilde \Psi\rangle$, where the two
emitted electrons propagate towards the same beamsplitter, only
contributes to the physical observables of interest to higher order in
tunnel barrier transparency and can be neglected. In what follows
below, we thus focus on the state $|\Psi\rangle$, describing one
electron propagating towards beamsplitter A and one towards B.

To clearly visualize the orbital entanglement, we present the state
$|\Psi\rangle$ in a first quantization notation. Introducing
$|1A,E\rangle|\uparrow\rangle$ for the operator
$c_{1A}^{\uparrow\dagger}(E)$, the properly symmetrized wavefunction
$|\Psi\rangle$ is given by
\begin{eqnarray}
|\Psi\rangle&=&\int_{-eV}^{eV}
dE\left(|1A,E\rangle|1B,-E\rangle+|1B,-E\rangle|1A,E\rangle+|2A,E\rangle|2B,-E\rangle+|2B,-E\rangle|2A,E\rangle\right)
\nonumber \\
&&\otimes\left(|\uparrow\rangle |\downarrow\rangle-|\downarrow\rangle |\uparrow\rangle\right).
\label{state0}
\end{eqnarray}
Here the first and second ket correspond to the first and second
particle respectively. We note that the total wavefunction is a direct
product of an antisymmetric (the spin singlet state of the Cooperpair)
spin wavefunction and a symmetric orbital wavefunctions. The orbital
entanglement with respect to emission of the pair across barrier 1 or
2 can be even more explicitly seen by writing the wave function as 
\begin{eqnarray}
|\Psi\rangle&=&|\Psi_{12}\rangle \otimes|\Psi_{AB}\rangle,
 \hspace{5mm}|\Psi_{12}\rangle=\left(|11\rangle+|22\rangle\right)/\sqrt{2},
 \nonumber \\ &&|\Psi_{AB}\rangle=\int_{-eV}^{eV} dE
 \left(|A,E\rangle|B,-E\rangle+|B,-E\rangle|A,E\rangle\right)\otimes\left(|\uparrow\rangle |\downarrow\rangle-|\downarrow\rangle |\uparrow\rangle\right)
\label{state}
\end{eqnarray}
The state $|\Psi_{12}\rangle$ is an orbitally entangled state with
respect to emission across barrier $1$ and $2$. This two-dimensional
space, called $12$-space below, plays the role of a pseudo-spin
space. The state $|\Psi_{AB}\rangle$ contains all additional
information, such as energy and spin dependence.

Propagating from the superconductors towards the beamsplitters the
electrons pick up phases $\varphi_{1A},\varphi_{1B},\varphi_{2A}$ and
$\varphi_{2B}$, where the index $1A$ etc. denotes the arm of the
normal conductor, see Fig. \ref{fig1}. These phases can be taken
energy independent for the system parameters considered. The
beamsplitters are assumed to have an adjustable transparency and to
support only one propagating mode. Such beamsplitters have been
realized in recent experiments \cite{Liu,Liu2}. The states $|+A
\rangle$ and $|-A\rangle$ ($|+B \rangle$ and $|-B\rangle$) for
electrons going out into the normal reservoirs and the states
$|1A\rangle$ and $|2A \rangle$ ( $|1B\rangle$ and $|2B \rangle$) of
the electrons incident on the beasmsplitters are related via energy
independent scattering matrices:
\begin{eqnarray}
\left[\begin{array}{c} |+A\rangle \\ |-A \rangle \end{array}\right]=
\left(\begin{array}{cc} \cos \phi_{A} & -\sin
\phi_{A} \\ \sin \phi_{A} & \cos
\phi_{A} \end{array}\right) \left[\begin{array}{c}
|2A \rangle \\ |1A \rangle \end{array}\right], \hspace{1cm} 
\left[\begin{array}{c} |+B\rangle \\ |-B \rangle \end{array}\right]=
\left(\begin{array}{cc} \cos \phi_{B} & -\sin
\phi_{B} \\ \sin \phi_{B} & \cos
\phi_{B} \end{array}\right) \left[\begin{array}{c}
|2B \rangle \\ |1B \rangle \end{array}\right]
\label{smat}
\end{eqnarray}
Possible scattering phases of the beamsplitters are incorporated in
the phases $\varphi_{j\eta}$, with $j=1,2$ and $\eta=A,B$. Under these
conditions the beamsplitters only affect the state in $12$-space,
i.e. $|\Psi_{12}\rangle$. On the normal reservoir side of the
beamsplitters, the state $|\Psi_{12}\rangle$ can be written
\begin{eqnarray}
|\Psi_{12}\rangle=c_{++}|++\rangle+c_{--}|--\rangle+c_{-+}|-+\rangle+c_{+-}|+-\rangle
\label{state2}
\end{eqnarray}
with the coefficients
\begin{eqnarray}
c_{++}&=&\left[\sin(\phi_A)\sin(\phi_B)e^{i(\varphi_{1A}+\varphi_{1B})}+\cos(\phi_B)\cos(\phi_A)e^{i(\varphi_{2A}+\varphi_{2B})}\right]/\sqrt{2},\nonumber
\\
c_{--}&=&\left[\cos(\phi_A)\cos(\phi_B)e^{i(\varphi_{1A}+\varphi_{1B})}+\sin(\phi_B)\sin(\phi_A)e^{i(\varphi_{2A}+\varphi_{2B})}\right]/\sqrt{2},
\nonumber \\
c_{-+}&=&\left[\cos(\phi_A)\sin(\phi_B)e^{i(\varphi_{1A}+\varphi_{1B})}-\cos(\phi_B)\sin(\phi_A)e^{i(\varphi_{2A}+\varphi_{2B})}\right]/\sqrt{2},\nonumber
\\
c_{+-}&=&\left[\sin(\phi_A)\cos(\phi_B)e^{i(\varphi_{1A}+\varphi_{1B})}-\sin(\phi_B)\cos(\phi_A)e^{i(\varphi_{2A}+\varphi_{2B})}\right]/\sqrt{2}.
\label{state3}
\end{eqnarray}
This provides a full description of the state of the particles exiting
into the reservoirs.

\section{Detection via violation of a Bell Inequality.}

The entanglement gives rise to a nonlocal correlation between the two
electrons, i.e. the outcome of a measurement at $A$ (e.g. electron in
A+) effects the probabilities for the outcome of the measurement in B
(electron in B+ or B-). The implications of such non-local
correlations have been at the heart of the discussion in quantum
mechanics for almost a century \cite{EPR,Bohr,Bohm}. Bell was the
first to show \cite{Bell} that the predictions of quantum mechanics
were incompatible with a whole class of theories, trying to explain
the non-local correlations with local, realistic arguments in the
spirit of Einstein, Podolsky and Rosen\ \cite{EPR}. Bells proof was
formulated in terms of an inequality which could not be violated by
any system described by a local, realistic theory. Systems described
by quantum mechanics could however violate such inequalities, showing
that local realistic theories were untenable.

Several approaches to Bell Inequalities (BI) in mesoscopic conductors
have been discussed \cite{Mait,Kaw,Cht,fazio}. Here we follow the
approach in Ref. \cite{Sam1}, which is closely related to Bells
original derivation. Starting from the two particle state emitted from
the superconductors, using the formulation in Eq. (\ref{state2}), we
can directly calculate the probabilities for joint detection of an
electron in reservoirs $\alpha A$ and one in $\beta B$
($\alpha,\beta=+,-$), given by the simple relation
\begin{equation}
P_{\alpha\beta}=|c_{\alpha\beta}|^2
\label{corrsimp}
\end{equation}
This gives directly
\begin{eqnarray}
P_{++}=P_{--}&=&\left[\cos^2(\phi_A)\cos^2(\phi_B)+\sin^2(\phi_B)\sin^2(\phi_A)\right.
\nonumber \\
&+& \left. \cos(2[\varphi_A-\varphi_B])\cos(\phi_A)\sin(\phi_B)\cos(\phi_B)\sin(\phi_A)\right]/2\nonumber \\  
P_{+-}=P_{-+}&=&\left[\cos^2(\phi_A)\sin^2(\phi_B)+\cos^2(\phi_B)\sin^2(\phi_A)\right.
\nonumber \\ 
&-& \left.
\cos(2[\varphi_A-\varphi_B])\cos(\phi_A)\sin(\phi_B)\cos(\phi_B)\sin(\phi_A)\right]/2
\label{corrs}
\end{eqnarray}
where we have introduced the phases
$2\varphi_A=\varphi_{1A}-\varphi_{2A}$ and
$2\varphi_B=\varphi_{2B}-\varphi_{1B}$. The phases $\varphi_A$ and
$\varphi_B$ can be varied independently and locally, at beamsplitters
A and B, by e.g. electrostatic sidegates changing the length of the
normal arms. Note that the joint detection probabilities are
normalized such that $P_{++}+P_{--}+P_{+-}+P_{-+}=1$.  Following Bell
and later formulations \cite{Clauser,Aspect}, we introduce the
correlation function
\begin{equation}
E({\bf a},{\bf b})=P_{++}({\bf a},{\bf b})-P_{+-}({\bf a},{\bf b})-P_{-+}({\bf a},{\bf b})+P_{--}({\bf a},{\bf b}).
\label{BI2}
\end{equation}
where ${\bf a}$ (${\bf b}$) collectively denotes the settings of the
phase $\phi_A$ and angle $\theta_A$ (phase $\phi_B$ and angle
$\theta_B$). In terms of these correlation functions for four
different measurements settings, denoted ${\bf a,a',b}$ and ${\bf
b'}$, the Bell Inequality can be formulated as
\begin{eqnarray}
-2\leq S_B\leq 2, \hspace{1cm} S_B\equiv E({\bf a},{\bf b})-E({\bf
 a},{\bf b'})+E({\bf a'},{\bf b})+E({\bf a'},{\bf b'}).
\label{BI}
\end{eqnarray}
where $S_B$ is the Bell parameter. This provides us with a BI in terms
of the joint detection probabilities $P_{\alpha\beta}$. However, in
mesoscopic systems, in contrast to optics, these quantities are not
experimentally accesible, they would demand a short-time coincidence
measurement. The natural observables are instead zero-frequency
current correlators,
\begin{eqnarray}
S_{\alpha\beta}\equiv \int_{-\infty}^{\infty}dt\langle \{\delta \hat
I_{\alpha A}(t) \delta \hat I_{\beta
B}(0)+\delta \hat I_{\beta B}(0)\delta \hat I_{\alpha A}(t) \}\rangle
\label{pendep}
\end{eqnarray}
where $\delta \hat I_{\alpha\eta}(t)=\hat I_{\alpha\eta}(t)-\langle
\hat I_{\alpha\eta}\rangle$ is the fluctuating part of the electrical
current $\hat I_{\alpha\eta}(t)$ in reservoir $\alpha\eta$. The
zero-frequency current correlators are calculated along the lines of
scattering theory \cite{Butt} and are given by \cite{Sam1} 
\begin{eqnarray}
S_{++}=S_{--}&=&\frac{4e^2}{h}|eV|P_{++}=\frac{4e^2}{h}|eV|P_{--}
\nonumber \\ 
S_{+-}=S_{-+}&=&\frac{4e^2}{h}|eV|P_{+-}=\frac{4e^2}{h}|eV|P_{-+}
\label{corrs2}
\end{eqnarray}
where $P_{\alpha\beta}$ are just the joint detection probabilites of
Eq. (\ref{corrs}). Thus, the zero-frequency current correlators are
directly proportional to the joint detection probabilites. The simple
result can be understood by considering the properties of the
time-dependent correlator $\langle \delta \hat I_{\alpha A}(t) \delta
\hat I_{\beta B}(0)\rangle$. It decays as $(\tau_c/t)^2$ for times
$t>\tau_c$, where $\tau_c=\hbar/eV$ is the correlation time of the
emitted pair. In the tunneling limit under consideration, $\Gamma \ll
1$, the correlation time is thus much smaller than the average time
between the arrival of two pairs $e/I \sim \hbar/eV\Gamma^2$. As a
result, only the two electrons within a pair are correlated with each
other, while electrons in different pairs are completely
uncorrelated. Thus, the zero frequency current correlator in
Eq. (\ref{pendep}) is just a coincidence counting measurement running
over a long time, collecting statistics over a large number of
pairs. This leads to the important result that the Bell inequality,
Eq.  (\ref{BI}), can be directly formulated in terms of the
zero-frequency current correlators in Eq.\ (\ref{pendep}), i.e. the
correlation function of Eq. (\ref{BI2}) is given by
\begin{equation}
E({\bf a},{\bf b})=\left[S_{++}({\bf a},{\bf b})-S_{+-}({\bf a},{\bf b})-S_{-+}({\bf a},{\bf b})+S_{--}({\bf a},{\bf b})\right]/P_0
\label{BIzf}
\end{equation}
where $P_0=(4e^2/h)|eV|$ is the proportionality constant between the
current correlator and the joint detection probability. 

It is important to note that BI's in electrical conductors quite
generally have to be investigated under conditions different from the
ones in optics. In optics, the discussion has to large extent been
focused on the detection problem, the fact that only a fraction of the
photons emitted are actually detected \cite{Clauser,Pearle,CH,Gisin3}
and how this effects the interpretation of an experimental violation
of a BI. This is in general not a problem in electrical conductors,
all electrons are detected in the reservoirs and contribute to the
statistics. There are however different aspects of BI's to be
considered in electrical conductors. First, the entagled quantities
are not free electrons but quasiparticle excitations out of the Fermi
sea. Second, there are several ``loopholes'', due to possible
interaction with phonons or other electrons or coupling to the
electromagnetic environment. All these phenomena open up the
possibility for local realistic theories that might explain the
observed phenomena. In addition, more sophisticated local realistic
theories allowing for some communication (possibly at the speed of
light) between the detectors or the source and the detectors, the so
called locality loophole, can hardly be ruled out in any mesoscopic
experiment (but have in fact been done so experimentally \cite{Weihs}
in optics).

With these limitations in mind, a violation of a BI in an electrical
conductor would neverthess provide a first indication of the nonlocal
properties of correlations between electrons. More importantly, the BI
treated completely within the framework of quantum mechanics can be
used as a tool to extract the amount of entanglement. This becomes
clear in the discussion below of the effect of dephasing on
entanglement. To investigate the parameter dependence of the
correlation functions in Eq. (\ref{BIzf}) we insert the expressions
for the zero frequency correlators in Eq. (\ref{corrs2}) and obtain
\begin{equation}
E({\bf a},{\bf b})=\cos (2\phi_A)\cos
(2\phi_B)+\cos(2[\varphi_A-\varphi_B])\sin (2\phi_A)\sin (2\phi_B).
\label{eimp}
\end{equation}
We note that there are two types of parameters that govern these
correlation functions, the phases $\varphi_{A}$ and $\varphi_B$ and
the angles $\phi_A$ and $\phi_B$. Putting the angles $\phi_{A}$ and
$\phi_B$ to $\pi/4$ (a semi-transparent beam-splitter) gives rise to
the correlation function in terms of the phases $\varphi_{A}$ and
$\varphi_{B}$
\begin{equation}
E(\varphi_A,\varphi_B)=\cos(2[\varphi_A-\varphi_B]).
\label{eimp2}
\end{equation}
Second, we can instead put the phases $2[\varphi_A-\varphi_B]$ to a
multiple of $2\pi$, giving a correlation function in terms of the
beam-splitter angles
\begin{equation}
E(\phi_A,\phi_B)=\cos (2\phi_A)\cos
(2\phi_B)+\sin(2\phi_A)\sin (2\phi_B)=\cos(2[\phi_A-\phi_B]).
\label{eimp3}
\end{equation}
This has exactly the same form as the correlation function in
Eq. (\ref{eimp2}). Inserting these correlation functions into the Bell
parameter in Eq. (\ref{BI}), and choosing e.g.
$\phi_A=\pi/8,\phi_B=\pi/4,\phi_A'=3\pi/8$ and $\phi_B'=\pi/2$ (or
equivalently the same values for the phases $\varphi_{j\eta}$) we get
a Bell parameter $S=2\sqrt{2}$. This gives a maximal violation of the
BI in Eq. (\ref{BI}).

From this, one gets the impression that it is of no importance what
parameters are adjusted. However, as investigated in detail below, the
two approaches lead to very different results for the ability to
violate the BI in the presence of dephasing.

\section{Effect of dephasing.}

There are many sources of dephasing in electrical
conductors. Dephasing can arise due to direct interaction between
different electrons or between electrons and phonons. Another source
is coupling of the electrons to the electromagnetic environment. An
investigation of dephasing in mesoscopic conductors is given by
Seelig, Pilgram and B\"uttiker in this issue. Focusing on systems with
geometries and physical properties similar to the one in
Fig. \ref{fig1}, we note that the effect of dephasing on the phase
dependence of the conductance in Mach-Zehnder interferometers was
recently studied in Refs. \cite{Seelig1,Seelig2}. Very recently, the
effect of dephasing on the phase dependence of the conductance as well
as the shot noise in Mach-Zender interferometers was investigated
experimentally \cite{ji}. Subsequently, a theory for the effect of
dephasing on the shot noise in the same geometry was proposed
\cite{marq}.  We also note that the effect of dephasing on the
possibility to violate a BI has been studied early in optics
\cite{ReidWalls}.

In this paper we study a simple model of the dephasing that
nevertheless illustrates the general effects of dephasing on the
entanglement and the possibility to violate the BI. The dephasing is
modelled phenomenologically via a density matrix
\begin{equation}
\rho=[|11\rangle\langle 11|+|22\rangle \langle
22|+\gamma(|11\rangle\langle 22|+|11\rangle \langle 22|)]/2
\label{densmat}
\end{equation}
which only affects the entanglement in $12$-space. The dephasing
parameter $\gamma$ can vary from $1$, when the system is in a pure,
coherent, fully entangled state and no dephasing is present, to $0$,
when the system is in a completely mixed state, fully dephased and
with the entanglement suppressed. We note that this density matrix
would e.g. arise from fluctuating phases $\varphi_{j\eta}$, discussed
in Refs. \cite{Seelig1,Seelig2,marq}. The density matrix in
Eq. (\ref{densmat}) gives rise to the correlation function, a
modification of Eq. (\ref{eimp}),
\begin{equation}
E({\bf a},{\bf b})=\cos (2\phi_A)\cos
(2\phi_B)+\gamma\cos(2[\varphi_A-\varphi_B])\sin (2\phi_A)\sin
(2\phi_B).
\label{eimp4}
\end{equation}
From this expression it is clear that the two approaches described
above to adjust the parameters of the correlation function give very
different results when trying to violate the BI. Fixing the angles
$\phi_A$ and $\phi_B$, the Eq. (\ref{eimp4}) becomes
\begin{equation}
E(\varphi_A,\varphi_B)=\gamma\cos(2[\varphi_A-\varphi_B])
\label{eimp5}
\end{equation}
Thus, the dephasing parameter $\gamma$ takes on the role of what is known in
the optical litterature as the visibility. For a visibility smaller
than $1/\sqrt{2}$, the BI can not be violated, which thus puts the
rather strict condition
\begin{equation}
\gamma>\frac{1}{\sqrt{2}}
\end{equation}
on the amount of dephasing allowed. Choosing the other approach above,
fixing the phases $\varphi_A$ and $\varphi_B$, we find
Eq. (\ref{eimp4}) modified to
\begin{equation}
E(\phi_A,\phi_B)=\cos (2\phi_A)\cos (2\phi_B)+\gamma\sin (2\phi_A)\sin
(2\phi_B).
\label{eimp6}
\end{equation}
As was pointed out in Ref. \cite{Sam1}, this allows in principle for a
violation of the BI for arbitrary dephasing. In the next section, this
is investigated in detail.

\section{Optimal violation.}

\begin{figure}[h]
\includegraphics[scale=0.29]{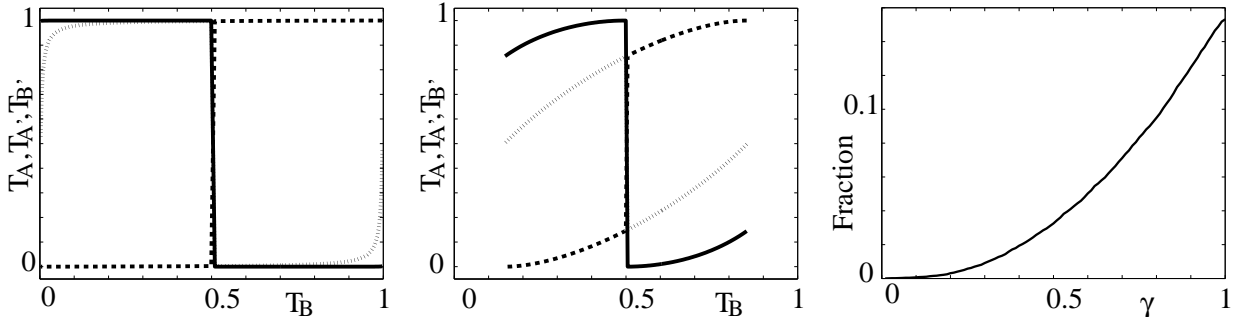}
\caption{Left and middle figures, from Ref. \cite{Sam1}: The
transmission probabilities $T_{A}$ (dashed),$T_A'$ (dotted) and $T_B'$
(solid) as a function of $T_{B}$ [$T_{\eta}=\cos^2(\phi_{\eta})]$,
giving optimal violation of the Bell inequalities for dephasing
parameters $\gamma=0.1$ (left) and $\gamma=1$. (middle). Right figure:
The fraction of the transmission probability space allowing for a
violation, as a function of the dephasing parameter $\gamma$.}
\label{andreev}
\end{figure}

We first try to find the angles which maximize the Bell parameter in
Eq. (\ref{BI}). By maximizing the Bell parameter with respect to first
$\phi_A$ and $\phi_{A}'$ and then to $\phi_B-\phi_{B}'$, we arrive at
the relations for the angles, all in the first quadrant
\begin{eqnarray}
\tan(2\phi_A)&=&-\gamma\cot(\phi_S), \nonumber \\
\tan(2\phi_{A}')&=&\gamma\tan(\phi_S), \nonumber \\
\tan(\phi_B-\phi_{B}')&=&\mbox{sign}[\cos(2\phi_A)]\sqrt{\frac{\tan^2(\phi_S)+\gamma^2}{\gamma^2\tan^2(\phi_S)+1}}
\end{eqnarray}
where $\phi_S=\phi_B+\phi_{B}'$ can be chosen at will. Inserting these
angles into the Bell parameter, we arrive at
\begin{equation}
S_B=2\sqrt{1+\gamma^2}
\label{BI3}
\end{equation}
showing that for an optimal choice of the angles
$\phi_A,\phi_{A}',\phi_B$ and $\phi_{B}'$, the BI can be violated for
arbitrarily strong dephasing, $\gamma>0$. The corresponding optimal
transmission probabilities $T_{\eta}=\cos^2(\phi_{\eta})$ are shown
for $\gamma=1$ and $\gamma=0.1$ in Fig.\ \ref{andreev}. 
The BI can thus in principle be violated for any amount of
dephasing. However it might be difficult to produce beam splitters
which can reach all transmission probabilities between $0$ and
$1$. This is not a serious problem in the absence of dephasing,
$\gamma=1$, a violation can be obtained for a large, order of unity,
fraction of the ``transmission probability space'' (see
Fig. \ref{andreev}). However, in the limit of strong dephasing, $\gamma
\ll 1$, the set of transmission probabilities for optimal violation
contains transmissions close to both $0$ and $1$, see
Fig. \ref{andreev}. Expecting unity transmission to be most
complicated to reach experimentally, we note that by instead choosing
transmission probabilities $T_A=T_B=0$, $T_B'=1/2$ and $T_A' \ll
\gamma$, the inequality in Eq. (\ref{BI}) becomes $2|1+\gamma
T_A'|\leq 2$. This gives a violation, although not maximal, for all
$\gamma\ll 1$.

Apart from dephasing there are several other effects such as
additional scattering phases, impurity scattering or asymmetric normal
conductor arms that might influence the ability to violate the
BI. These effects are discussed in Ref. \cite{Sam1}. Another
possibility is that, due to asymmetries of the tunnel barriers
$\Gamma_1 \neq \Gamma_2$, the amplitude for the process where the pair
is emitted to $|11\rangle$ is different from the process where it is
emitted to $|22\rangle$. This gives rise to a state, in 12-space,
\begin{equation}
|\Psi_{12}\rangle=\frac{1}{\sqrt{\Gamma_1^2+\Gamma_2^2}}\left(\Gamma_1|11\rangle+\Gamma_2|22\rangle\right).
\end{equation}
In this case \cite{Gisin91,Abour} we directly get the same expression
for the coherence function in Eq. (\ref{eimp4}), with an effective
\begin{equation}
\gamma=\frac{2\Gamma_1\Gamma_2}{\Gamma_1^2+\Gamma_2^2}.
\end{equation}
This shows that it is in principle possible to violate BI for
arbitrary asymmetry as well.

\section{Conclusions}

As a conclusion, we have investigated orbital entanglement in
mesoscopic conductors with the focus on the effect of dephasing.  The
entanglement is detected via violation of a Bell Inequality formulated
in terms of zero-frequency current correlations. We have investigated
how the dephasing affects the possibility to violate the Bell
Inequality and how system parameters can be adjusted for optimal
violation.

\section*{Acknowledgements} 

This work was supported by the Swiss National Science Foundation and
the Swiss network for Materials with Novel Electronic Properties.

\end{document}